\documentclass[amssymb,amsmath,prl,twocolumn,floatfix,showpacs]{revtex4}
\usepackage{graphicx}

\begin{document}
\title{Quantum information transport to multiple receivers.}

\author{Andrew D.~Greentree}
\affiliation{Centre for Quantum Computer Technology, School of
Physics, The University of Melbourne, Melbourne, Victoria 3010,
Australia.}

\author{Simon J.~Devitt}
\affiliation{Centre for Quantum Computer Technology, School of
Physics, The University of Melbourne, Melbourne, Victoria 3010,
Australia.}

\author{Lloyd C.~L.~Hollenberg}
\affiliation{Centre for Quantum Computer Technology, School of
Physics, The University of Melbourne, Melbourne, Victoria 3010,
Australia.}

\date{\today}

\begin{abstract}
The importance of transporting quantum information and entanglement
with high fidelity cannot be overemphasized.  We present a scheme
based on adiabatic passage that allows for transportation of a
qubit, operator measurements and entanglement, using a 1-D array of
quantum sites with a single sender (Alice) and multiple receivers
(Bobs). Alice need not know which Bob is the receiver, and if
several Bobs try to receive the signal, they obtain a superposition
state which can be used to realize two-qubit operator measurements
for the generation of maximally entangled states.
\end{abstract}


\maketitle


Allied to the efforts to build a working quantum computer (QC) is
the requirement to replicate, in a quantum framework, the necessary
features of a classical computer.  In particular, some kind of
quantum bus would be highly advantageous to allow a form of
distributed QC \cite{bib:EisertPRA2000}. Long range quantum
information transfer usually exploits teleportation
\cite{bib:CopseyIEEE2003}, or flying qubits
\cite{bib:DiVincenzoFPhys2000}; we consider a mechanism more related
to a quantum wire or fanout. Fanout operations are forbidden quantum
mechanically, as they necessarily imply cloning, however,
considering the closest quantum analog leads to a new approach to
quantum communication, described here.

A na\"{i}ve approach to transport in quantum system is via
sequential swap gates between sites. This approach is often
undesirable due to, for example, noise introduced by sensitive
nonadiabatic controls, poor level of gate control, insufficient
bandwidth or impractical gate density \cite{bib:CopseyIEEE2003}.
Many authors have begun to examine alternatives
\cite{bib:Trans,bib:ChristandlPRL2004,bib:Haselgrove2004}
considering schemes where a desired coupling is set up (usually
statically) and the system allowed to evolve until the information
transfer has occured.  In such schemes the receivers may be passive
\cite{bib:ChristandlPRL2004}, or active \cite{bib:Haselgrove2004},
but it is usually assumed that neither sender nor receiver can
modify the qubit chain, except for control of their own qubit or
qubits and local coupling to the chain.

We propose an extremely general alternative for adiabatic transfer
of a particle between positional quantum states.  An obvious
application of this is as a transport mechanism for ion trap QCs. In
one approach \cite{bib:KeilpinskiNature2002} a scheme for
transporting ions sequentially from storage zones to interaction
sites was proposed via a microtrap array \cite{bib:CiracNature2000}:
the Quantum Charge Coupled Device.  With minor modification, our
scheme provides an interesting alternative. One attractive feature
of the present scheme is that requirements on quantum state guidance
are minimized, and sympathetic cooling following transport should
not be required. One could also consider solid-state realizations of
this scheme in a patterned GaAs quantum dot array
\cite{bib:GaAs,bib:HayashiPRL} or where the confining potentials are
realized using ionized P donors in a Si matrix
\cite{bib:Kane,bib:HollenbergPRB,bib:Hollenberg:Scalable}.

\begin{figure}[tb!]
\includegraphics[width=0.8\columnwidth,clip]{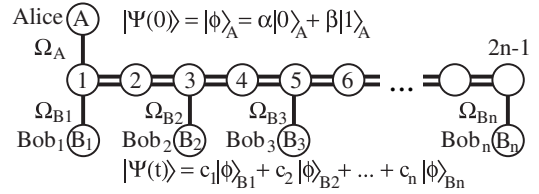}
\caption{\label{fig:Alice} Quantum bus for qubit transfer from Alice
to the Bobs. The initial state $|\Psi(0)\rangle = |\phi\rangle_A$
goes to a spatial superposition state $|\Psi(t)\rangle = \sum_j c_j
|\phi\rangle_j$. The bus is defined by tunnel coupled sites 1 to
$n$. Static, high tunneling matrix elements (TME) are shown by
double lines, lower, controlled TME by single lines. Alice controls
one site, $A$, and the Bobs are represented by the sites $B_i$,
connected to the bus at sites $1, 3 \cdots 2n-1$. Alice controls
$\Omega_A$, and Bob$_j$ controls $\Omega_{B_j}$.}
\end{figure}

We consider a quasi-one-dimensional chain of quantum sites,
realized, by the empty or singly occupied states of a positional
eigenstate, see Fig.~\ref{fig:Alice}. The sender (Alice) distributes
information via a qubit using the chain to a series of receivers
(Bob$_1$ to Bob$_n$).  Alice and each Bob control one site each,
attached to the central chain, and Alice need not know who receives
the signal.

Our scheme differs from previously discussed schemes by the use of
spatial adiabatic passage to transport the qubit. The addition of
multiple receivers and its inherent flexibility is a significant
departure from previous adiabatic protocols
\cite{bib:STIRAP,bib:CTAP,bib:Eckert2004,bib:SiewertASSP2004}, and
we term this new transport scheme MRAP - Multiple Receiver Adiabatic
Passage. In addition, the previous adiabatic passage protocols are
further augmented by transporting an extra, spectator degree of
freedom in addition to the spatial degree of freedom. This allows
flexible information transfer, and even distributed entanglement in
an adiabatic context. The geometry is chosen to ensure that there is
always an odd number of quantum sites between Alice and each Bob,
see Fig.~\ref{fig:Alice}, which is required by the adiabatic passage
protocol \cite{bib:CTAP}, analogous versions of the three-site
protocol have been considered in optical lattices
\cite{bib:Eckert2004} and Cooper-Pair boxes
\cite{bib:SiewertASSP2004}, and a two-electron, three-site variant
has been suggested for entanglement distillation
\cite{bib:Fabian0412229}.  Although we explicitly consider spatial
transfer of particles, information could be transferred using spin
chains (e.g. that of Ref. \onlinecite{bib:ChristandlPRL2004}) with
time-varying adiabatic coupling sequences instead of static
couplings.

To investigate transport, we write the Hamiltonian for a single
qubit carried by a particle in a positional array
\begin{eqnarray}
\mathcal{H} = \sum_{\sigma = 0,1} \left[ \sum_{i=1}^{2n-1}
    \left(\frac{E_{i,\sigma}}{2} c^{\dag}_{i,\sigma} c_{i,\sigma} +
    \Omega_S c^{\dag}_{i+1,\sigma} c_{i,\sigma} \right) \right. \nonumber \\
+  \left( \frac{E_A}{2} c^{\dag}_{A,\sigma} c_{A,\sigma} + \Omega_A c^{\dag}_{1,\sigma} c_{A,\sigma} \right)\nonumber \\
+ \left. \sum_{j = 1}^{n} \left( \frac{E_{B_j}}{2}
c^{\dag}_{B_j,\sigma} c_{B_j,\sigma} + \Omega_{B_j}
c^{\dag}_{B_j,\sigma} c_{2j+1,\sigma} \right)\right] +
\textrm{h.c.}, \label{eq:Ham}
\end{eqnarray}
where we have introduced the (externally controlled) site energies,
$E$, and $\Omega_S$ is the tunneling matrix element (TME) along the
bus, which is not varied during the protocol, $\Omega_A$ is the TME
between $A$ and $1$ which Alice can control, whilst $\Omega_{B_j}$
is the TME between $B_j$ and $2j-1$, and $c_{i,\sigma}$ is the
annihilation operator for a qubit with state $\sigma=0,1$ on site
$i$ for $i=A,1 \cdots 2n-1,B_1 \cdots B_n$.  Control of these TMEs
is by varying the potential barrier between the sites and the chain.
The exact method for this variation is implementation dependent, but
for a GaAs or P:Si system could be via surface gates
\cite{bib:HayashiPRL,bib:HollenbergPRB}, or mean well separation in
an optical lattice \cite{bib:Eckert2004}. For notational brevity we
do not indicate unoccupied sites explicitly, so that
$c^\dagger_{i,\sigma}|\mathrm{vac}\rangle = |\sigma\rangle_i$.
Eq.~\ref{eq:Ham} comprises three terms, the first corresponds to the
energy of the particle in the sites on the chain, and the TMEs
between chain sites, the second to the energy of the particle on
Alice's site, and the coupling from Alice to the chain, and the
final term to the energy of the particle at the Bobs' sites and
their tunneling to the chain. For $n$ Bobs there must be at least
$2n-1$ sites in the chain, so we assume this number (extra sites in
the chain do not interfere with the scheme as discussesd below). As
the qubit degree of freedom is decoupled from the positional degree
of freedom, it is carried along as a `spectator' storing
information, but otherwise unaffected by the transfer.

\begin{figure}[tb!]
\includegraphics[width=0.8\columnwidth,clip]{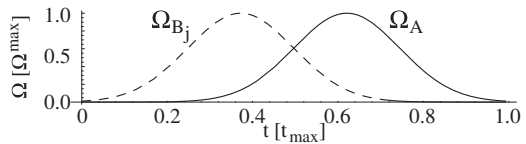}
\caption{\label{fig:Pulse} Counter-intuitive sequence for transport
from Alice to Bobs. Each receiver Bob chooses the same coherent
tunneling profile.  On this scale, the TME between bus sites would
be at least 10. Increasing the length of time for the protocol
increases the fidelity, up to the limits allowed by dephasing.}
\end{figure}

To realize the counter-intuitive pulse sequence, we set all of the
site energies to 0 (i.e. $E_{i,\sigma} = 0$ for $i = 1 \cdots 2n-1,
A, B_1 \cdots B_n$, using the external control. The TMEs are
modulated (again via external control) in a Gaussian fashion
according to (see Fig.~\ref{fig:Pulse})
\begin{eqnarray}
\Omega_{A}(t) = \Omega^{\max} \exp \left\{ -(\left[t-(t_{\max}/2 + s)^2 \right]/(2 s^2) \right\}, \nonumber \\
\Omega_{B_j}(t) = \Omega^{\max}_{B_j} \exp \left\{
-(\left[t-(t_{\max}/2 - s)^2 \right]/(2 s^2) \right\},
\end{eqnarray}
where $\Omega^{\max} \ll \Omega_S$, and $s$ is the width of the
applied pulses. $\Omega^{\max}_{B_j} = \Omega^{\max}$ if Bob$_{j}$
wishes to receive a signal from Alice, and $\Omega^{\max}_{B_j} = 0$
otherwise.  The scheme is extremely robust to the choice of
modulation, and in common with conventional adiabatic transfer
schemes alternatives to Gaussians have little effect providing the
adiabaticity criterion is satisfied \cite{bib:STIRAP}. As no control
of the chain is possible apart from modifications of the TMEs in the
vicinity of Alice and Bob, $\Omega_S$ is constant.

The first case to consider is a chain where only the final Bob
chooses to receive the qubit from Alice, i.e. $\Omega_{B_i} = 0, i =
1 \cdots n-1$.  In this case the MRAP protocol reduces to previous
adiabatic protocols \cite{bib:CTAP} and will not be treated here.
More generally, however, we show that the protocol works for
\emph{any} Bob, or indeed for multiple Bobs \emph{simultaneously}:
quantum fanout.

Alice broadcasts her qubit, and the Bobs have colluded so there is
only one receiver, Bob$_j$, which is not communicated to Alice.
Remarkably, no extension to the above scheme is required. When both
Alice and Bob$_j$ perform MRAP, Bob$_j$ receives the information in
a time-scale defined by the \emph{total} length of the bus, and
\emph{independent} of the site where Bob$_j$ is situated. In marked
departure from most previous schemes, the information does not
propagate along the bus. Rather, one eigenstate is smoothly
transformed from being located at Alice's site, to Bob$_j$'s site,
without occupying the bus at any stage. To understand this, consider
the two-Bob protocol (with three chain sites), with Alice and
Bob$_1$ connected to site 1, and Bob$_2$ connected to site 3.
Adiabatic transport utilizes the eigenstates  of Eq.~\ref{eq:Ham}
which have zero eigenvalue, i.e. the two-dimensional null space of
$\mathcal{H}$, spanned by
\begin{eqnarray}
|\psi_1\rangle = \left(\Omega_{B_1}|\phi\rangle_A -
\Omega_A |\phi\rangle_{B_1}\right)\sqrt{\Omega_A^2 + \Omega_{B_1}^2}, \nonumber \\
|\psi_2\rangle = \frac{\Omega_{B_2}|\phi\rangle_A -
\frac{\Omega_A\Omega_{B_2}}{\Omega_S} |\phi\rangle_2 + \Omega_A
|\phi\rangle_{B_2}}{\sqrt{\Omega_A^2 + \Omega_{B_2}^2}},
\end{eqnarray}
where $\Omega_S \gg \Omega_A, \Omega_{B_1}, \Omega_{B_2}$, and
$|\phi\rangle_i \equiv \alpha|0\rangle_i + \beta|1\rangle_i$. Any
linear combination of these two states is also a null state, so it
suffices to find the state adiabatically connected to Alice's site
at $t=0$, i.e. $|\Psi(t=0)\rangle = |\phi\rangle_{A}$. If
$\Omega_{B_2}(t) = 0 \forall t$, $|\psi_1\rangle$ is adiabatically
connected to $|\Psi(t=0)\rangle$ and the qubit is transferred from
$|\phi\rangle_A$ to $|\phi\rangle_{B_1}$, if $\Omega_{B_1}(t) = 0
\forall t$, then $|\psi_2\rangle$ is adiabatically connected to
$|\phi\rangle_A$, and the qubit transferred from $|\phi\rangle_A$ to
$|\phi\rangle_{B_2}$. Hence the qubit can be sent from Alice to
either Bob, without Alice knowing which Bob is the receiver.

If both Bobs are receivers, they choose $\Omega_{B_1}(t) =
\Omega_{B_2}(t)$ and $(|\psi_1\rangle - |\psi_2\rangle)/2$ is
adiabatically connected to $|\phi\rangle_A$. The final state of the
system after MRAP is $(|\phi\rangle_{B_1} -
|\phi\rangle_{B_2})/\sqrt{2}$, which is quantum fanout, with both
Bobs sharing an equal positional superposition of the qubit.  We
stress that have not cloned Alice's qubit, and measurements of the
qubit position will collapse it at either $B_1$ or $B_2$.  For
adiabatic transport, the adiabaticity criterion \cite{bib:STIRAP}
must be satisfied, i.e., the inverse transfer time must be small
compared with the energy gap between states, $E = [(\Omega_A^2 +
\Omega_{B_1}^2 + \Omega_{B_2}^2)/2]^{1/2}$ for large $\Omega_S$.

In the general case the null space is spanned by
\begin{multline}
|\psi_j\rangle =  \\
\frac{\Omega_{B_j}|\phi\rangle_{A}
  + \sum_{k = 1}^{j-1} \frac{\Omega_A\Omega_{B_j}}{(-1)^k \Omega_S}|\phi\rangle_{2k}
  + (-1)^{j} \Omega_A |\phi\rangle_{B_j}}{\sqrt{\Omega_A^2 + \Omega_{B_j}^2}},
\end{multline}
and up to known signs, all receiver Bobs obtain an equal
superposition of the qubit.  One can show that the energy gap
between the zero and next nearest eigenstate is $E_{\mathrm{gap}} =
[(\Omega_A^2 + \sum_{k = 1}^{j}\Omega_{B_k}^2)/j]^{1/2}$, for the
$j-$Bob protocol, so the scaling for more Bobs goes as
$[(1+j)/j]^{1/2}$.

We have performed preliminary studies of the robustness of this
scheme, and find a similar resistance to errors as other adiabatic
protocols. If the simultaneity of the Bob pulses is not exact, or if
the $\Omega_B$ pulse areas are not exactly the same, then the
superposition state shared following the protocol will not be exact.
Providing adiabaticity is satisfied, such errors introduce a
monotonic decrease in fidelity.  It is difficult to explore the full
state space because of the large range of parameters, however, the
point is that MRAP is not exponentially sensitive to errors. We have
also solved the case where the chain is cyclic, i.e. where site 1 is
connected to sites 2 and $2n$, and the protocol works without
modification. These results will be presented in more detail
elsewhere.

We have shown the most general case of Alice sending a qubit to the
Bobs, but a special case would be where Alice could be a factory of
pure states.  MRAP could be used to send these states with high
fidelity around a quantum network, important for many QC
architectures.

The only difference between Alice and Bobs is the order in which
they vary their coupling to the bus. Therefore the Bobs can also
perform inter-Bob communication by assuming the role of Alice or Bob
as required.  Reversing the protocol for the two-Bob case (with
$\Omega_{B_1}=\Omega_{B_2}$) gives the transformations
\begin{eqnarray}
|\phi\rangle_{B_1}\Rightarrow (1/\sqrt{2}|\phi\rangle_A +
(1/2)\left(|\phi\rangle_{B_1} + |\phi\rangle_{B_2}\right), \\
|\phi\rangle_{B_2}\Rightarrow -(1/\sqrt{2})|\phi\rangle_A +
(1/2)\left(|\phi\rangle_{B_1} + |\phi\rangle_{B_2}\right).
\end{eqnarray}

We have not included the effects of dephasing here, as it will vary
significantly between different implementations.  In most practical
systems of particle transfer, one expects particle localization to
dominate over spontaneous emission, i.e. $T_2 \ll T_1$.  Hence we
will ignore $T_1$.  However, $T_2$ processes will take the system
out of the null space, and will be detrimental to the transfer
protocol. Dephasing has been considered analytically in STIRAP
\cite{bib:Dephasing} and numerically for higher order protocols
\cite{bib:Hollenberg:Scalable}. If the minimum transfer time is
satisfied, the transfer failure probability goes as $\Gamma_2
T_{\mathrm{tot}}$ where $\Gamma_2$ is the dephasing rate and
$T_{\mathrm{tot}}$ is the \emph{total} protocol time, and this
requirement is no different from the requirements for charge-qubit
systems. Our numerical models show that MRAP is not inherently more
sensitive than the 1-D spatial adiabatic passage, although the total
time that superposition states must be maintained will be longer
than in the 1-D case, with corresponding (linear) decrease of
robustness.

Considering implementations, the total time for the protocol must be
$T_{\mathrm{tot}} \gtrsim 10 /\Omega_{\max}$ \cite{bib:CTAP}, and
the chain TMEs must be $\Omega_S \gtrsim 10 \Omega_{\max}$.  In a
P:Si system\cite{bib:Hollenberg:Scalable}, with Alice and Bob site
separations from the chain of $30\mathrm{nm}$, and interchain
separations of $20\mathrm{nm}$, give rise to $\Omega_{\max} \sim 100
\mathrm{GHz}$ and $\Omega_S \sim 1 \mathrm{THz}$, which gives a
probability of transfer error of $10^{-2}$ for $\Gamma_2 = 100
\mathrm{MHz}$ for $T_{\mathrm{tot}} \sim 2 \mathrm{ns}$, which is
certainly feasible (though unmeasured) given current projections of
P:Si architectures. Hu \textit{et al.} \cite{bib:HuPRB2005} suggest
that GaAs quantum dots with TMEs in the same range could be achieved
with inter-dot spacings of $30-35 \mathrm{nm}$.  Petta \textit{et
al.} \cite{bib:PettaPRL2004} measured charge dephasing rates of
$\Gamma_2 \sim 10 \mathrm{GHz}$, suggesting that a proof of
principle demonstration of MRAP is already possible, although
improvements in $\Gamma_2$ are needed before a practical GaAs
implementation is possible. Ion trap and optical lattice systems,
however, show the most promise for demonstrations. Eckert \textit{et
al.} \cite{bib:Eckert2004} estimate adiabatic timescales for the
three-state protocol of order milliseconds, and because the transfer
is in vacuum, $\Gamma_2$ should be small compared with this rate.

\begin{figure}[tb!]
\includegraphics[width=0.8\columnwidth,clip]{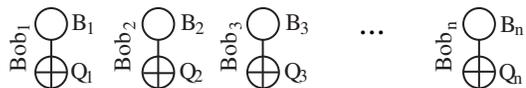}
\caption{\label{fig:B-Q} Each Bob has, in addition to their site,
$B_j$, a qubit, $Q_j$, and a CNOT or CZ gate with $B_j$ as control
and $Q_j$ as target.  This allows MRAP to be used for operator
measurements and entanglement between the $Q_j$.}
\end{figure}

MRAP can be extended to realize two-qubit operator measurements
\cite{bib:ChuangBook}.  The extension requires augmenting the
existing protocol with extra qubits and the ability to perform
entanglement operations between the Bob sites and these qubits. By
using the bus as a mediator for entanglement, our protocol has a
similar aim to the multisplitter of Paternostro \textit{et al.}
\cite{bib:PaternostroPRL2005}, but arises from a very different
mechanism. In addition to their vacant sites, $B_j$, each Bob has a
qubit, $Q_j$, and can perform either a CNOT or CZ operation between
sites $B_j$ (control) and $Q_j$ (target), depicted in
Fig.~\ref{fig:B-Q}.  An an example, we show the protocol for two-Bob
MRAP, where the multi-Bob bus forms an effective qutrit ancilla
(formed by the states $|1\rangle_A, |1\rangle_{B_1},
|1\rangle_{B_2}$), and we demonstrate projective measurements of
$U_1 U_2\equiv \sigma_{U,Q_1} \otimes \sigma_{U,Q_2}$ for $U = X,Z$.

1. Initially, the $Q_j$'s are in the arbitrary state
$|\Phi\rangle_{Q_1,Q_2}$, and the bus in state $|1\rangle_{A}$. The
total system state is
\begin{eqnarray*}
|\Psi\rangle = |1\rangle_A |\Phi\rangle_{Q_1,Q_2}.
\end{eqnarray*}

2. MRAP is performed, the system's state becomes
\begin{eqnarray*}
|\Psi\rangle = (1/\sqrt{2})\left(|1\rangle_{B_1} -
|1\rangle_{B_2}\right)|\Phi\rangle_{Q_1,Q_2}.
\end{eqnarray*}

3. The Bobs perform a Controlled-$U$ operation between sites $B_i$
and $Q_i$, where the action of the controlled operation is trivial
when $B_i$ is unoccupied, the state evolves to
\begin{eqnarray*}
|\Psi\rangle = (1/\sqrt{2})\left(|1\rangle_{B_1}U_{Q_1} I_{Q_2} -
|1\rangle_{B_2} I_{Q_1} U_{Q_2}\right)|\Phi\rangle_{Q_1,Q_2},
\end{eqnarray*}
where $I$ is the identity operator.

4. MRAP transfer is reversed, generating the state
\begin{eqnarray*}
|\Psi\rangle = (1/2)|1\rangle_A \left(U_{Q_1}
I_{Q_2} + I_{Q_1} U_{Q_2}\right) |\Phi\rangle_{Q_1,Q_2} + \\
(2\sqrt{2})^{-1}\left(|1\rangle_{B_1} + |1\rangle_{B_2}\right)
\left(U_{Q_1} I_{Q_2} - I_{Q_1} U_{Q_2}\right)
|\Phi\rangle_{Q_1,Q_2}.
\end{eqnarray*}

5. A measurement is performed at Alice, detecting the bus qubit with
probability $1/4$, and projecting the state of $[Q_1,Q_2]$ to
$\left(U_{Q_1} I_{Q_2} + I_{Q_1} U_{Q_2}\right)
|\Phi\rangle_{Q_1,Q_2}$, i.e. the $+1$ eigenstate of $U_{Q_1}
U_{Q_2}$, if successful.

6. If no qubit was measured at Alice, the system is projected to
$\left(U_{Q_1} I_{Q_2} - I_{Q_1} U_{Q_2}\right)
|\Phi\rangle_{Q_1,Q_2}$, which is the $-1$ eigenstate of
$U_{Q_1}U_{Q_2}$, so a $\sigma_z$ at B$_2$ allows the qubit to be
deterministically returned to Alice in another reverse of the MRAP
protocol.  Hence MRAP affords a complete two-qubit operator
measurement of $XX$ and $ZZ$.

As the above protocol gives $XX$ and $ZZ$ operator measurements on
physically separated qubits, we may use this to create
multi-particle stabilizer states, e.g. the $N$ particle GHZ state
$|GHZ\rangle_N = (1/2)^{N/2}(|00\cdots0\rangle_N + |11\cdots
1\rangle_N)$ (for convenience we choose $N$ even). First, initialize
$Q_1$ to $Q_N$ to $|0\rangle$. Then, perform $X_{2i-1}X_{2i}$
stabilizer measurements via two-Bob MRAP on the pairs of $Q_{2i-1}$
and $Q_{2i}$ for $i = 1 ... N/2$, creating a series of independent
two-particle Bell states $\bigotimes_i (|10\rangle \pm
|01\rangle)_{Q_{2i-1}Q_{2i}}$ where the relative sign is known from
the projective measurement result at Alice. Local single qubit
operations can be used to convert the Bell pairs to $|00\rangle +
|11\rangle$. Next $Z_{2i}Z_{2i+1}$ stabilizer measurements are
performed between the Bell pairs, and this is sufficient to project
the computer into $|GHZ\rangle_N$ (up to local operations).  If one
has access to multiple Alice's, and the ability to break the chain
freely (with surface gates), then these operations can be performed
in parallel, meaning that the $|GHZ\rangle_N$ state can be formed in
two MRAP steps.

We have introduced a transport mechanism for quantum information
around a network, based on adiabatic passage.  With minor
modification our scheme can also be used for generating entanglement
and two-qubit measurements. The scheme is ideally suited as an
alternative to conventional ionic transport in an ion trap quantum
computer.  However its utility not restricted to ion traps, and it
should have wide applicability to all architectures, especially
solid-state quantum computing architectures.

ADG thanks Fujitsu for support whilst at University of Cambridge,
and discussions with S. G. Schirmer, D. K. L. Oi, J.~H.~Cole, A. G.
Fowqler and C.~J.~Wellard. LCLH was supported by the Alexander von
Humboldt Foundation, and thanks the van Delft group at LMU for their
hospitality, and discussions with F. Wilhelm. This work was
supported by the Australian Research Council, US National Security
Agency (NSA), Advanced Research and Development Activity (ARDA) and
Army Research Office (ARO) under contract W911NF-04-1-0290.


\end{document}